\documentclass{aa}
\usepackage[dvips]{graphicx}
\usepackage{psfig}
\usepackage{ltxtable}
\def\AMM{NH$_3$}
\def\NTHP{N$_{2}$H$^{+}$}

\def\CEIO{\mbox{C$^{18}$O}}

\def\WAT{H$_2$O}

\def\MOLH{H$_2$}
\def\MOLN{N$_{2}$}
\def\MOLO{O$_{2}$}

\def\kms{\mbox{km~s$^{-1}$}}
\def\percc{cm$^{-3}$}

\def\cmsq{cm$^{-2}$}

\def\mic{\mbox{$\mu$m}}
\def\arcsec{$^{''}$}
\def\arcmin{$^{'}$}

\def\lesssim{\mathrel{\hbox{\rlap{\hbox{\lower4pt\hbox{$\sim$}}}\hbox{$<$}}}}
\def\gtrsim{\mathrel{\hbox{\rlap{\hbox{\lower4pt\hbox{$\sim$}}}\hbox{$>$}}}}
\begin{document}
\title{The  depletion of NO in pre--protostellar cores}


\author{M.~Akyilmaz \inst{1} \and D.R. Flower\inst{1}
\and P. Hily--Blant\inst{2}
\and G. Pineau des For\^{e}ts\inst{3,4}
\and C.M. Walmsley\inst{5}}

\institute{Physics Department, The University, Durham DH1 3LE, UK
\and IRAM, 300 Rue de la Piscine, F-38406 Saint Martin d'H\`{e}res, France
\and IAS (UMR 8617 du CNRS), Universit\'{e} de Paris--Sud, F-91405 Orsay, France
\and LERMA (UMR 8112 du CNRS), Observatoire de Paris, 61 Avenue de l'Observatoire, F-75014, Paris, France
\and INAF, Osservatorio Astrofisico di Arcetri, Largo Enrico Fermi 5, I-50125 Firenze, Italy}

\offprints{C.M. Walmsley}


\abstract%
	{Understanding the depletion of heavy elements is a
	  fundamental step towards determining the structure of
	  pre--protostellar cores just prior to collapse.}%
	{We study the dependence of the NO abundance on position
	  in the pre--protostellar cores L1544 and L183.}%
	{We observed the 150~GHz and 250~GHz transitions of NO
	  and the 93~GHz transitions of \NTHP \ towards L1544
	  and L183 using the IRAM 30~m telescope. We compare the
	  variation of the NO column density with position in
	  these objects with the H column density derived from
	  dust emission measurements.}%
	{We find that NO behaves differently from \NTHP \ and
	  appears to be partially depleted in the high density
	  core of L1544. Other oxygen--containing compounds are
	  also likely to be partially depleted in dense--core
	  nuclei.  The principal conclusions are that: the
	  prestellar core L1544 is likely to be `carbon--rich';
	  the nitrogen chemistry did not reach equilibrium prior
	  to gravitational collapse, {\bf and nitrogen is
	  initially (at densities of the order of
	  $10^4$~cm$^{-3}$) mainly in atomic form}; the grain
	  sticking probabilities of atomic C, N and, probably, O
	  are significantly smaller than unity.}

\titlerunning{Depletion of NO}
\maketitle

%

\section{Introduction}

It is now accepted that most, if not all, C--containing
species deplete on to dust grain surfaces in the central,
high--density regions of prestellar cores. Thus, when one
compares a map of C$^{18}$O, for example, with that of the
dust emission, one sees that the CO isotopomer traces only
an outer shell and not the high--density interior ($n_{\rm
H} \gtrsim 3\times 10^4$ cm$^{-3}$)\footnote{In this paper,
we express gas densities implicitly in terms of the total
number density of protons, $n_{\rm H} \equiv n({\rm H}) +
2n({\rm H}_2) + n({\rm H}^+) + ... \approx 2n({\rm
H}_2)$. However, to be consistent with some of the cited
publications, we quote explicitly values of $n$(H$_2$) from
time to time.}; this limits our understanding of prestellar
cores, because the molecular line emission provides
information on the kinematics.

Surprisingly, N--containing species (in particular, \AMM \
and \NTHP) manage to remain in the gas phase up to densities
of a few times $10^5$ cm$^{-3}$. It is not entirely clear
what happens at still higher densities, but it seems likely
that the N--containing species disappear from the gas phase
when $n_{\rm H} \gtrsim 10^6$ cm$^{-3}$ (see, for example,
Belloche \& Andr\'{e} 2004, Pagani et al. 2005).  The
different behaviour of the N--containing species was
originally attributed to the relatively high volatility of
N$_{2}$ on grain surfaces (\cite{bergin97},
\cite{aikawa01}), but recent laboratory measurements
(\cite{oberg05},~\cite{bisschop06}) show that N$_{2}$ is
only marginally more volatile than CO.  This result throws
doubt on much recent modelling of the chemistry of
prestellar cores and poses the question of whether another
more volatile and abundant form of nitrogen is responsible
for the survival of N--containing species to high density.
Another possibility is that atomic nitrogen does not stick
effectively to grain surfaces. In recent studies (Flower et
al. 2005, 2006), we have explored this suggestion and found
that the best fit of the models to the observed abundances
of \AMM \ and \NTHP \ is obtained when the sticking
probabilities of both atomic N and O are low and,
additionally, the mean grain surface area per H atom is a
factor 5--10 lower than in the diffuse interstellar medium;
this implies that considerable grain growth has occurred
prior to the prestellar--core phase.

  In all of this discussion, the abundance of
{\it oxygen--containing species}, such as OH and water, has been somewhat neglected, and for good  
reason. OH and water are hard to observe with reasonable
(arc min or better) resolution in cold clouds
(see, for example, Goldsmith \& Li 2005, Bergin \& Snell 2002). Thus we  
do not
presently know whether oxygen--containing species survive in the
gas phase in the regions where \NTHP \  and \AMM \ are observed. In the
long run, observations of water with the HIFI instrument on the  
Herschel satellite or of OH
with the Square Kilometer Array (SKA) may provide an answer. In the meantime, we need to consider  
alternative observational strategies, using chemical models of  
prestellar cores as a guide. If atomic oxygen sticks to grain surfaces no more efficiently than atomic nitrogen, OH and \WAT \  might be expected to survive in the gas phase similarly  
to \AMM \ and N$_2$H$^+$. 

  An important aspect of this problem involves NO, which is believed to be a key molecule
in the nitrogen chemistry. NO mediates the transformation of
  atomic into molecular nitrogen, in the reactions

\begin{equation}
{\rm N} + {\rm OH} \rightarrow {\rm H} + {\rm NO}
\label{equ1}
\end{equation}
and 

\begin{equation}
{\rm N} + {\rm NO} \rightarrow {\rm O} + {\rm N}_{2}
\label{equ2}
\end{equation}
Atomic and molecular nitrogen are expected to be the major
forms of elemental nitrogen in the gas phase.  Under
conditions in which reactions~(\ref{equ1}) and (\ref{equ2})
dominate the formation and destruction, respectively, of NO,
the ratio of the abundance of NO to that of OH is given
simply by the ratio of the rate coefficients for these
reactions, $k_1/k_2$; $k_1/k_2 \approx 1$, providing that
there are no energy barriers to these reactions, which might
reduce differentially the values of their rate coefficients
at low temperatures. Then, one might consider NO to be a
proxy for OH, providing information on O--containing species
in the gas phase. OH forms through a sequence of reactions
which is initiated by O(H$_3^+$, H$_2$)OH$^+$ (and the
analogous reactions with the deuterated isotopes, H$_2$D$^+$
and D$_2$H$^+$), followed by hydrogenation by H$_2$ and
dissociative recombination with electrons. OH is destroyed
principally in the reactions O(OH, O$_2$)H and N(OH,
NO)H. Thus, the abundance of OH is tied to the abundances of
H$_3^+$, H$_2$D$^+$ and D$_2$H$^+$, and OH is a minor
oxygen--containing species.

NO is a known interstellar molecule, which has been detected
in L183 with a fractional abundance of around $10^{-7}$ (see
Gerin et al. 1992, 1993). The value quoted by these authors
was an overestimate, as it was based on the \CEIO \ column
density, and CO is now known to be depleted in L183 (Pagani
et al. 2004, 2005). Nonetheless, the fractional abundance of
NO is high, which encouraged us to investigate the spatial
variation of the NO abundance in the cores L183 and L1544.
For these objects, estimates of the column densities of
molecular hydrogen, $N$(\MOLH), which are independent of the
degree of depletion, may be made from the existing maps of
the dust emission. Our objective was to compare the spatial
variations of the NO and H$_2$ (from the dust) column
densities.  In addition, we observed simultaneously with NO
the $J = 1 \rightarrow 0$ transitions of N$_2$H$^+$, which
are known to follow roughly the dust emission.

  In Sect.~2, we describe our observations and, in
Sect.~3, give the observational results. Comparisons with model  
predictions are made in Sect.~4, and our concluding remarks are to be  
found in Sect.~5.

\section{Observations}
\label{obs}

    The observations were carried out during two sessions, in July and
August 2005, using the IRAM 30~m telescope. We observed simultaneously
with the facility 3~mm, 2~mm, and 1.3~mm receivers. The 3~mm receiver was tuned to the
\NTHP \ $J = 1 \rightarrow 0$ 93.17632~GHz line, the 2~mm receiver to the centre of the NO $J = 3/2 \rightarrow 1/2$ $\Pi^{+}$ and $\Pi^{-}$ transitions at 150.36146~GHz,
and the 1.3~mm receiver to the centre of the NO $J = 5/2 \rightarrow 3/2$ $\Pi^{+}$ and $\Pi^{-}$ transitions  at 250.61664~GHz.  The
half--power beam widths at these three frequencies are 27\arcsec,  
16\arcsec, and 10\arcsec, respectively, and the corresponding  
main--beam efficiencies are
0.78, 0.68, and 0.48. Pointing was checked at intervals of roughly 2  
hours
and was found to be accurate within 2\arcsec.  We used central  
positions
(J2000) of R.A.~=~$05^{\rm h}04^{\rm m}16^{\rm s}.9$,  
Dec.~=~25$^{\circ}10^{'}47^{''}.7$ for L1544 and
  R.A.~=~$15^{\rm h}54^{\rm m}08^{\rm s}.8$,  
Dec.~=~$-02^{\circ}52^{'}44^{''}.0$
for L183, and the offsets specified below are with respect to these  
positions.

We used the facility SIS receivers in frequency--switching mode
with system temperatures of
150--180~K at 3~mm, 350--450~K at 2~mm, and 1000--2000~K at 1.3~mm.
  At 3~mm, we used  the
facility VESPA  autocorrelator as a spectral backend with 20~kHz
channel spacing over a 20~MHz band, allowing us to cover all
7 hyperfine satellites of the \NTHP($J = 1 \rightarrow 0$) transition.  At 2~mm  
and at 1.3~mm,
the NO lines are too widely separated to be covered in
one band with adequate resolution.  We therefore split VESPA into
6 parts of 20~MHz bandwidth and 20~kHz resolution at 2~mm
and 3 parts with 40~MHz bandwidth and 40~kHz resolution at 1.3~mm. The
different sections of the autocorrelator were offset so as to
cover the frequencies and transitions summarized in Table~\ref{nofreq}.
The corresponding velocity resolutions were 0.04~\kms \ at 2~mm and
0.05~\kms \ at 1.3~mm.  The data were processed using the GILDAS
package\footnote{See URL http://www.iram.fr/IRAMFR/GILDAS/}.

\begin{table}
\caption{\sc NO line parameters }
\vspace{1em}
\begin{tabular}{ccc}
\hline
\hline
Transition  & Frequency  & Line Strength$^{1}$ \\
$J^{'}, F^{'}\rightarrow J^{''}, F^{''}$   & GHz           &               
\\
\hline
$\Pi^{+}$  &   &  \\
$\frac {3}{2},\frac {5}{2} \rightarrow \frac {1}{2},\frac {3}{2}$ &   
150.17646  &  0.500 \\
$\frac {3}{2},\frac {3}{2} \rightarrow \frac {1}{2},\frac {1}{2}$  &  
150.19876   & 0.186 \\
$\frac {3}{2},\frac {3}{2} \rightarrow \frac {1}{2},\frac {3}{2}$ &  
150.21874   & 0.148 \\
$\frac {3}{2},\frac {1}{2} \rightarrow \frac {1}{2},\frac {1}{2}$ &  
150.22565  & 0.148 \\
$\Pi^{-}$  &  &  \\
$\frac {3}{2},\frac {5}{2} \rightarrow \frac {1}{2},\frac {3}{2}$ &  
150.54646  &  0.500  \\
$\frac {3}{2},\frac {1}{2} \rightarrow \frac {1}{2},\frac {1}{2}$  &  
150.58055 & 0.148 \\
$\Pi^{+}$  &  &  \\
$\frac {5}{2},\frac {7}{2} \rightarrow \frac {3}{2},\frac {5}{2}$ &  
250.43684  & 0.445 \\
$\Pi^{-}$  &  &  \\
$\frac {5}{2},\frac {7}{2} \rightarrow \frac {3}{2},\frac {5}{2}$ &  
250.79643  & 0.445 \\
$\frac {5}{2},\frac {5}{2} \rightarrow \frac {3}{2},\frac {3}{2}$ &  
250.81561  & 0.280 \\
\hline
\hline
\multicolumn{3}{l}{(1) Gerin et al.~(1992)}
\end{tabular}
\label{nofreq}
\end{table}

\section{Results}

In Appendix~A, we describe the method used to determine the column densities of \NTHP, which are used in the chemical analysis presented in the following Section. Here, we discuss the derivation of the column densities of NO for each of the two prestellar cores which we observed.

\subsection{L1544}
  In the time available, rather than attempting to
map the NO emission, we observed along two cuts, in the
NW--SE and NE--SW directions, centred on the origin given in Sect.~\ref{obs};
these directions correspond roughly to the major and minor axes, respectively, of  
the dust emission, which we assume to be a good representation of
the total  column density distribution.  Furthermore, the  intensities of the NO
150~GHz (2~mm) lines are expected to yield a good approximation to the true
NO column density distribution (see~\cite{gerin92}), as
long as the lines are optically thin; this is likely to be the
case, given their relative weakness (the  low $J$ levels are
expected to be thermalized in L1544). Furthermore, the relative intensities
of the different hyperfine components are observed to be
close to the values expected on the basis of the line strengths in  
Table~\ref{nofreq}.  Averaging the corresponding
transitions from the  $\Pi ^{+}$ and
$\Pi ^{-}$ bands,  we find a ratio of $0.46\pm 0.07$ 
between the $J = 3/2 \rightarrow 1/2$ $F = 3/2 \rightarrow 1/2$ and $F = 5/2 \rightarrow 3/2$ lines (at zero offset, (0,0): see Table~\ref{lpar}), as compared to 0.37 using
the line strengths in Table~\ref{nofreq}. 
Similarly, the ratio of the average of the three 2~mm lines with strength  
0.148
in Table~\ref{nofreq} to the two lines with strength 0.500
is measured to be $0.30\pm 0.04$, as compared with 0.30 in the optically  
thin LTE limit.  Thus, we find that the NO hyperfine satellite intensities  
are almost consistent, to within the error bars, with the low optical depth limit.
We can then convert the integrated intensity $I$(2~mm)
in the $J = 3/2 \rightarrow 1/2$ $F = 5/2 \rightarrow 3/2$ (average of  
$\Pi ^{+}$ and $\Pi ^{-}$) into an NO column density for a given  
excitation temperature, $T_{\rm {ex}}$(NO).

\begin{table}
\caption{NO transitions observed towards L1544 (0,0), with uncertainties in parentheses. The linewidths of the 150~GHz transitions were determined by means of a gaussian fit, using the CLASS procedure; the linewidths of the (noisier) 250 GHz lines were fixed to that of the 150.17646~GHz transition.}
\vspace{1em}
\begin{tabular}{cccc}
\hline
\hline
   Frequency  & Integrated intensity  & $V_{\rm {lsr}}$ & $\Delta  
v$ \\
   GHz           & K~\kms  & \kms    & \kms              \\
\hline
    150.17646  & 0.22(0.02)  & 7.18(0.01)   & 0.38(0.02)  \\
    150.19876  & 0.09(0.01)  & 7.14(0.02)   & 0.35(0.04)  \\
    150.21874  & 0.05(0.01)  & 7.12(0.02)   & 0.30(0.04)  \\
    150.22565  & 0.06(0.01)  & 7.12(0.02)   & 0.26(0.04)  \\
    150.54646  & 0.20(0.01)  & 7.11(0.01)   & 0.34(0.02)  \\
    150.58055  & 0.04(0.01)  & 7.13(0.03)   & 0.34(0.06)  \\
    250.43684  & 0.26(0.04)  & 7.28(0.05)   & 0.38 \\
    250.79643  & 0.22(0.04)  & 7.26(0.04)   & 0.38  \\
\hline
\hline
\end{tabular}
\label{lpar}
\end{table}

  Because NO has a small dipole moment, the critical density (at which the rates of collisional and radiative de-excitation are equal) is of order  
$10^4$ \percc, and it seems reasonable to suppose thermalization of the  
NO levels of interest to us here. Based on the \AMM \ results of~\cite{tafalla02}, we assume $T_{\rm {ex}} = 8.75$~K and find that the column density of NO cm$^{-2}$ in L1544 should be given approximately by 

\begin{displaymath}
N({\rm NO}) = 4.3\times 10^{15}\, I(2\,{\rm mm}) 
\end{displaymath}
where $I$(2~mm) K~km~s$^{-1}$ denotes the intensity of the NO 2~mm (150~GHz) lines (see~\cite{gerin92}, table~3). We note that new VLA observations by
Crapsi et al. (2005b) suggest that the temperature decreases
towards the dust peak; but, as we shall see below, NO is likely to be  
depleted at high densities, and so we use the  
single--dish temperature estimate.  An independent check of the NO  
excitation is provided
by our measurements of the $J = 5/2 \rightarrow 3/2$ 250~GHz lines.
 From Table~\ref{lpar}, we conclude that $I$(1.2~mm)/$I$(2~mm) = $1.14\pm 0.25$,
where, in addition to the formal errors  given in the Table, we have
assumed a 15~percent calibration error. Here, $I$(1.2~mm) is the average
of the integrated intensities of the two $J = 5/2 \rightarrow 3/2$ $F = 7/2 \rightarrow 5/2$ transitions. Using the formulation
of Gerin et al. (1992), we find that the corresponding excitation  
temperature between the $J = 5/2$ and $J = 3/2$ levels is $T_{\rm  
{ex}} = 13.5\pm 4$~K, which
is higher than the  single--dish \AMM \ measurement but consistent with it, to  
within the combined error bars.

\begin{figure}
  \centering
  \includegraphics[height=15cm]{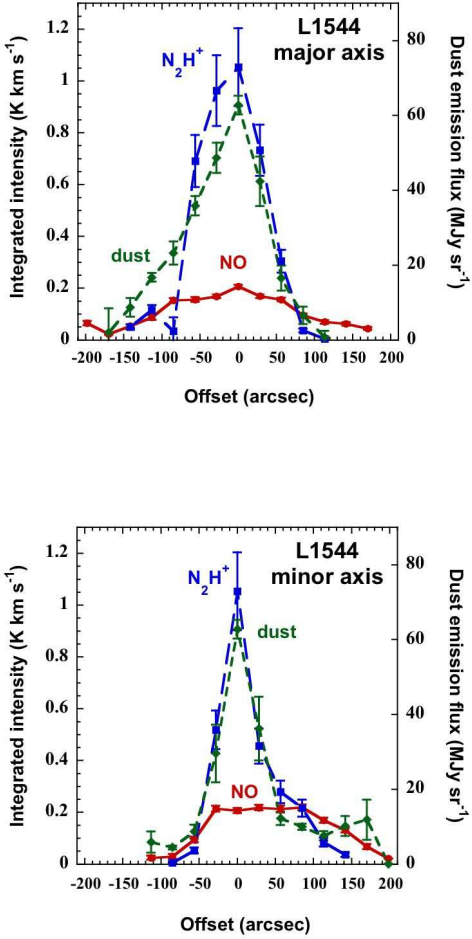}
  \caption{Plots of the average of the integrated
	intensities of the NO $\Pi^{+}$ and $\Pi^{-}$ $J, F =
	3/2, 5/2 \rightarrow J, F = 1/2, 3/2$ lines (see Table
	\ref{nofreq}), denoted by full curve joining circles
	with error bars, which have approximately the same size
	as the symbols, plotted against offset from the (0,0)
	position in L1544. Increasing offset is in the sense of
	increasing R.A. The top panel shows the cut along the
	NE--SW major axis, and the bottom panel the cut along
	the NW--SE minor axis. We show also for comparison the
	integrated intensity of the 93.17632~GHz transition of
	\NTHP \ (squares joined by long broken curve) and the
	1.2~mm dust emission (diamonds joined by short broken
	curve), from Ward-Thompson et al. (1999). The error bars
	plotted correspond to RMS deviations from the fits
	(gaussian for the lines); thus we neglect systematic
	errors arising from the calibration, for example. The NO
	measurements involved long integration times, and the
	two components ($\Pi ^{+}$ and $\Pi ^{-}$) were
	averaged, whence the small error bars in this case. The
	intensities derive from main--beam brightness
	temperatures.}
  \label{figcut}
\end{figure}

\begin{figure}
  \centering
  \includegraphics[height=10cm]{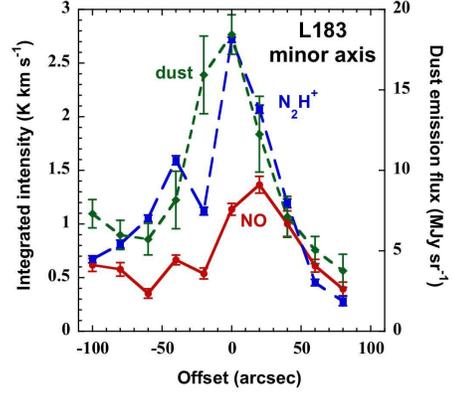}
  \caption{Plots of the average of the integrated
	intensities of the NO $\Pi^{+}$ and $\Pi^{-}$ $J, F =
	3/2, 5/2 \rightarrow J, F = 1/2, 3/2$ lines (see Table
	\ref{nofreq}), denoted by full curve joining circles
	with error bars, against offset from the (0,0) position
	in L183. Increasing offset is in the sense of increasing
	R.A. The cut is along the E--W (minor) axis.  We show
	also for comparison the integrated intensity of the
	93.17632~GHz transition of \NTHP \ (squares joined by
	long broken curve) and the 1.2~mm dust emission
	(diamonds joined by short broken curve), from Pagani et
	al. (2004). The intensities derive from main--beam
	brightness temperatures.}
  \label{figcutbis}
\end{figure}

In Fig.~\ref{figcut}, we show our observations of NO at 2~mm
for both the NE--SW and NW--SE cuts, together with the
corresponding measurements of the dust emission (from the
map of~\cite{wardt99}) and our 3~mm observations of \NTHP \
$J, F_1, F = 1, 0, 1 \rightarrow 0, 1, 2$ (93.17632~GHz)
transition.  In essence, these data show that, while the
dust and the \NTHP \ emission peak at the (0,0) position and
have an angular size of about 1\arcmin \ by 2\arcmin \
(0.04~pc by 0.08~pc at the adopted distance of 140~pc of
L1544; \cite{crapsi05a}), the NO emission extends over at
least 3\arcmin\ and weakens, relative to the \NTHP \ and
dust emission, towards the (0,0) position. We can quantify
this result by using estimates from Fig.~\ref{figcut} of the
flux in the dust continuum emission. We derive a peak of
emission at 1.2~mm of 225~mJy in the beam, which
corresponds, for an assumed dust absorption coefficient
$\kappa $ per unit mass density $\rho $ of gas, $\kappa
/\rho = 0.01$~cm$^{2}$~g$^{-1}$, to an \MOLH \ column
density of $1.2\times 10^{23}$~\cmsq ; we have assumed that
the dust temperature is equal to the excitation temperature,
$T_{\rm {ex}} = 8.75$~K.  This conversion factor was used to
derive the column density ratio, $x$(NO) =
$N$(NO)/$N$(\MOLH), shown in Fig.~\ref{noplot}.  We see that
the observed column density ratio is approximately $8\times
10^{-9}$ at the dust peak and increases to values of
$3\times 10^{-8}$ to $5\times 10^{-8}$ at offsets of around
1\arcmin \ (equivalent to 0.04~pc, or 8000~AU) along the
minor axis, where the \MOLH \ column density, derived from
the dust emission, is a factor of 6 lower than at the peak.
Along the major axis, the variation is less marked, but
there is also an increase in the column density ratio away
from the peak of the dust emission. We suggest in the next
Section that this behaviour is probably related to the
partial depletion, at high densities, of the main forms of
oxygen.

\subsection{L183}

Our observed line parameters, towards the (0,0) position of
L183, are given in Table~\ref{l183par}. In this case, only
one cut was possible in the time available, in an E--W
direction, and the results are shown in
Fig.~\ref{figcutbis}.

We have used a procedure to derive the NO abundance similar
to that for L1544. We assume a temperature of 9~K, based on
the \AMM \ results of Ungerechts et al. (1980), in order to
compute both the NO and \MOLH \ column densities.  We use
additionally the dust emission map of Pagani et al. (2004).
As discussed below, our NO column density peak in L183
appears to be shifted by about 10\arcsec \ in R.A., relative
to the dust peak (roughly 1000~AU for an assumed distance of
110~pc; \cite{franco89}).  We infer column density ratios,
$N$(NO)/$N$(\MOLH ), in the range $1\times 10^{-8}$ to
$3\times 10^{-8}$, similar to the values derived for L1544.

\begin{table}
\caption{NO transitions observed towards L183 (0,0), with uncertainties in parentheses}
\vspace{1em}
\begin{tabular}{cccc}
\hline
\hline
   Frequency  & Integrated intensity  & $V_{\rm {lsr}}$ & $\Delta v$ \\
   GHz           & K~\kms  & \kms    & \kms              \\
\hline
    150.17646  & 0.47(0.015)  & 2.37(0.01)   & 0.38(0.02)  \\
    150.19876  & 0.18(0.015)  & 2.37(0.02)   & 0.44(0.04)  \\
    150.21874  & 0.14(0.014)  & 2.29(0.02)   & 0.38(0.04)  \\
    150.22565  & 0.14(0.014)  & 2.35(0.02)   & 0.32(0.04)  \\
    150.54646  & 0.46(0.016)  & 2.38(0.01)   & 0.38(0.02)  \\
    150.58055  & 0.12(0.012)  & 2.38(0.02)   & 0.29(0.04)  \\
    250.43684  & 0.27(0.060)  & 2.33(0.01)   & 0.11(0.03)  \\
    250.79643  & 0.32(0.060)  & 2.38(0.02)   & 0.27(0.06)  \\
\hline
\hline
\end{tabular}
\label{l183par}
\end{table}

\section{Modelling the NO abundance distribution}

The simple models discussed below are based on the
assumption of spherical symmetry, which is clearly a rough
approximation for both of the prestellar cores which we
consider. Indeed, Doty et al. (2005) have modelled L1544 as
a prolate spheroid, with an axis ratio of approximately
2. Such departures from spherical symmetry imply that the
timescale for collapse is different along each of the
principal axes. Furthermore, determinations of the thermal
pressure in these objects and estimates of their magnetic
field strengths suggest that the timescale for collapse is
probably larger than for free--fall. Thus, the
``free--fall'' collapse model which we have adopted is
undoubtedly a zero--order approximation. Nevertheless, for
the purposes of comparing observations of different chemical
species, and hence deducing the chemical and physical
conditions in the cores, the present approach is probably
adequate. In any case, to go further would necessitate a
more complete knowledge of the dynamics of the cores.

\subsection{Empirical model}
\label{empirical}

\begin{figure}
  \centering
  \includegraphics[height=15cm]{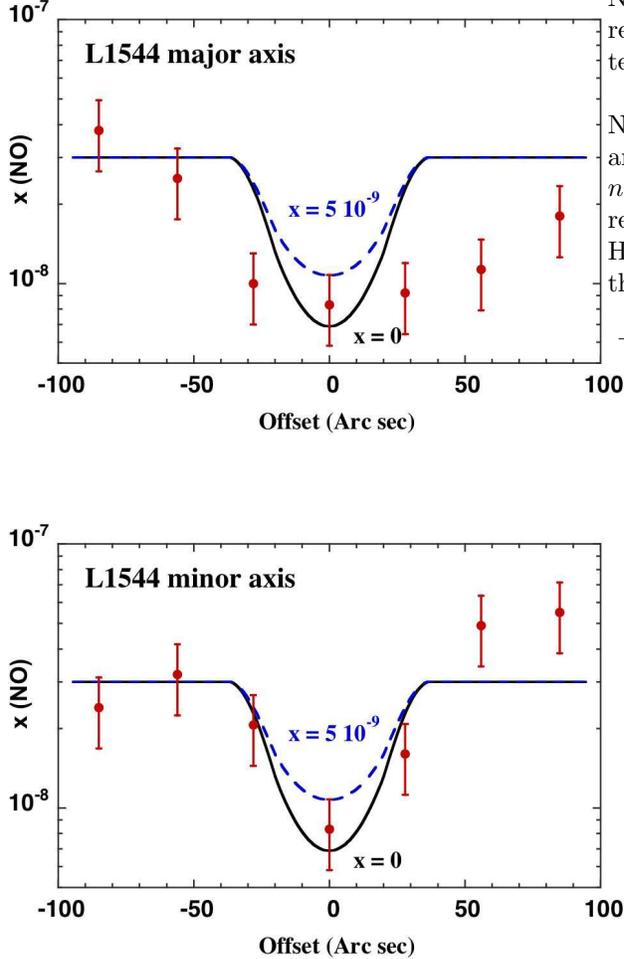}
  \caption{Observed and predicted NO to H$_2$ column density
	ratio, $x$(NO) = $N$(NO)/$N$(\MOLH ), for a model of
	L1544 in which the Tafalla et al. (2002) density
	distribution (cf. Sect.~\ref{f-f}) is adopted. Two
	different assumptions (indicated on the Figure) are made
	regarding the value of $n$(NO)/$n$(\MOLH ) in the
	central region, offsets $r \le 30$\arcsec \ (4000~AU):
	$n({\rm NO})/n({\rm H}_2) = 5\,10^{-9}$ (broken curve);
	$n({\rm NO})/n({\rm H}_2) = 0$ (full curve).  Outside
	this region, $n$(NO)/$n$(\MOLH ) = $3\, 10^{-8}$ is
	assumed. The filled circles with error bars denote the
	observed values.}
  \label{noplot}
\end{figure}

First, we simulate the observed NO column density
distribution, relative to the \MOLH \ column density deduced
from the dust emission (cf. Fig.~\ref{figcut}), by means of
a simple empirical model, according to which
$n$(NO)/$n$(\MOLH ) behaves as a step function, centred on
the (0,0) position. The ratio $x$(NO) = $N$(NO)/$N$(\MOLH )
is then computed, assuming spherical symmetry and convolving
with the telescope beam, taken to have a width of
17\arcsec. In Fig.~\ref{noplot}, we plot, for the case of
L1544, two cases, one in which $n$(NO)/$n$(\MOLH ) =
$5\times 10^{-9}$ for offsets $r \le 30$\arcsec \ (4000~AU),
and $n$(NO)/$n$(\MOLH ) = $3\times 10^{-8}$ for $r >
30$\arcsec \, and the other in which the ratio
$n$(NO)/$n$(\MOLH ) is assumed to be zero within the central
zone. It may be seen from Fig.~\ref{noplot} that the value
of $n$(NO)/$n$(\MOLH ) = $5\times 10^{-9}$ for the abundance
ratio in the central zone appears to be a reasonable upper
limit, given the error bars on the observations. The
asymmetries of the observed abundance ratio are not
reproduced by the model, which assumes spherical symmetry.

The central zone, in which $x$(NO) decreases, is smaller
than has been estimated for CO and some other species; the
decrease occurs at a density $n({\rm H}_2) \approx 4\times
10^5$ \percc.  Assuming an exponential decrease of the
fractional abundance with radial distance, Tafalla et
al. (2002) derived characteristic (1/e) densities of
$5.5\times 10^4$ \percc \ for \CEIO \ and $1.7\times 10^5$
\percc \ for CS. Thus, NO appears to be intermediate in its
depletion characteristics between the C--containing species,
which are strongly affected by CO depletion, and N--bearing
species like \NTHP \ and \AMM.

An analogous model for our R.A. cut in L183, in which the
Crapsi et al. (2005a) density distribution was adopted (a
central density of $10^6$ \percc \ and a power--law decrease
in the density beyond a radius of 4200~AU), shows the
displacement, to which we alluded above, of the observed
minimum of the NO abundance ratio from the dust peak. In
this case also, it will be necessary to map fully the NO
column density distribution to take the analysis further.

\subsection{Chemical considerations}
\label{chemical}

In prestellar cores, the nitrogen chemistry depends on
gas--phase conversion of atomic into molecular nitrogen.  As
was mentioned in the Introduction, the most efficient scheme
for this conversion is believed to involve NO as an
intermediate, through reactions~(\ref{equ1}) and
(\ref{equ2}) above. The ratio of the abundance of NO to that
of OH is then given by $k_1/k_2$, the ratio of the rate
coefficients for these reactions.

This relation no longer applies when the atomic nitrogen
abundance falls below the value for which the destruction of
NO by N, in reaction~(\ref{equ2}), has the same rate as
direct depletion of NO on to the grains, i.e. when $$n_{\rm
H}k_{\rm dep}({\rm NO}) = n({\rm N})k_2,$$ or $$x({\rm N})
\equiv n({\rm N})/n_{\rm H} = k_{\rm dep}({\rm NO})/k_2.$$
In these expressions, $$k_{\rm dep}({\rm NO}) = (n_{\rm
g}/n_{\rm H})\pi a_{\rm g}^2v_{\rm th},$$ where $n_{\rm g}$
is the grain number density, $a_{\rm g}$ is the grain
radius, and $v_{\rm th}$ is the thermal speed of the NO
molecules. Taking $T = 10$~K, we find $k_{\rm dep}({\rm NO})
= 1.4\times 10^{-18}(0.5/a_{\rm g})$, where $a_{\rm g}$ is
in $\mu $m and $k_{\rm dep}$ in cm$^3$~s$^{-1}$. With $k_2 =
3.5\times 10^{-11}$ cm$^3$~s$^{-1}$ (Baulch et al.~2005), we
find that, when the fractional abundance of atomic nitrogen
exceeds $x({\rm N}) = 4\times 10^{-8}$, gas--phase
destruction of NO by N dominates, whereas, below this value,
NO is removed predominantly by freeze--out on to the
grains. A similar argument can be applied to OH, and so the
abundances of NO and OH remain related in both regimes.

The rate coefficients for both of the reactions~(\ref{equ1})
and (\ref{equ2}) must be sufficiently large at low
temperatures to ensure the conversion of N into N$_2$, the
precursor of N$_2$H$^+$ and NH$_3$; but the measurement of
rate coefficients for neutral--neutral reactions at very low
temperatures is demanding (Sims 2005), and so the values of
$k_1$ and $k_2$ remain uncertain. As will be seen in
Sect.~\ref{f-f}, our observations of NO and \NTHP \ support
the existence of a small barrier to reaction~(\ref{equ2}),
which inhibits the destruction of NO by N at temperatures as
low as $T = 10$~K.

We recall that our observations indicate that, unlike \AMM \
and \NTHP, NO does not show a constant or even an increasing
abundance, relative to H$_2$, for densities $n_{\rm H}
\gtrsim 3\times 10^5$ \percc. Instead, NO appears to
decrease in relative abundance approaching the dust emission
peak. Here we review some possible chemical explanations of
this discrepancy.

\begin{itemize}

\item OH may be under--abundant in the high density regions
  of at least some prestellar cores.  This situation could
  arise for the same reasons that \MOLO \ and water are much
  less abundant than expected in prestellar cores (see, for
  example, \cite{pagani03}, \cite{snell00}).  Surplus oxygen
  -- that is, oxygen which is not bound in CO -- may have
  frozen out prior to the start of the collapse, giving rise
  to a situation whereby, once CO has also frozen out, there
  is more N than O in the gas phase; this assumes that N
  does {\it not} freeze out simultaneously with CO. In the
  lower density regions, in which CO is still present in the
  gas phase, the C/O elemental abundance ratio would be
  close to unity, and the chemistry would have
  ``carbon--rich'' characteristics, with, for example, large
  abundances of cyanopolyynes and similar species. In this
  case, the main source of atomic oxygen is the reaction of
  CO with He$^+$, which is produced by cosmic ray ionization
  of He, and hence the fractional abundances of both OH and
  NO are dependent on the cosmic ray flux. Also, their
  abundances are determined to some extent by how rapidly
  atomic oxygen sticks to the grains, an issue which will be
  discussed in Sect.~\ref{f-f} below.

\item There may be chemical pathways from N to \MOLN \ which
  we (and others) have overlooked.  Any conversion of N into
  N$^{+}$, through charge exchange with He$^{+}$, for
  example, would lead to the production of NH and other
  hydrides and the subsequent formation of \MOLN \ in
  neutral--neutral reactions.  However, trial calculations
  which we have carried out suggest that this particular
  route is not likely to be significant. More promising is
  the possibility that, under conditions in which the
  gas--phase abundance of elemental carbon approaches that
  of elemental oxygen, the reactions
\begin{equation}
  {\rm N} + {\rm CH} \rightarrow {\rm H} + {\rm CN}
  \label{equ3}
\end{equation}
and 

\begin{equation}
{\rm N} + {\rm CN} \rightarrow {\rm C} + {\rm N}_{2}
\label{equ4}
\end{equation}
assume importance relative to reactions~(\ref{equ1}) and
(\ref{equ2}). We consider this possibility further in
Sect.~\ref{f-f}.

\end{itemize}

We conclude that the main forms of oxygen (like the main
forms of carbon) appear to have frozen on to dust--grain
surfaces at densities approaching $10^6$ \percc \ and,
moreover, that some form of nitrogen is more volatile and
hence more abundant in the gas phase than any form of
oxygen. Another conclusion is that, in the densest parts of
prestellar cores, oxygen exists mainly as solid H$_2$O, CO,
and perhaps CO$_{2}$, with a fractional abundance in the
solid state which is larger than the values derived for
lower density ($\lesssim 3\times 10^4$ \percc )
material. Having eliminated other possibilities, we arrive
at the proposal that atomic nitrogen does not stick
effectively to icy dust grains, even at kinetic temperatures
close to 10~K (although we note that a small abundance of OH
or CH is also necessary to the formation of N$_2$).  This
proposal is {\bf not in conflict} with the recent laboratory
studies of Bisschop et al. (2006) and \"{O}berg et
al. (2005), which suggest that \MOLN \ freezes out similarly
to CO.  Then, atomic N is likely to be the main repository
of nitrogen in the gas phase.

\subsection{Free--fall collapse model}
\label{f-f}

\begin{figure*}
  \centering
  \includegraphics[height=15cm]{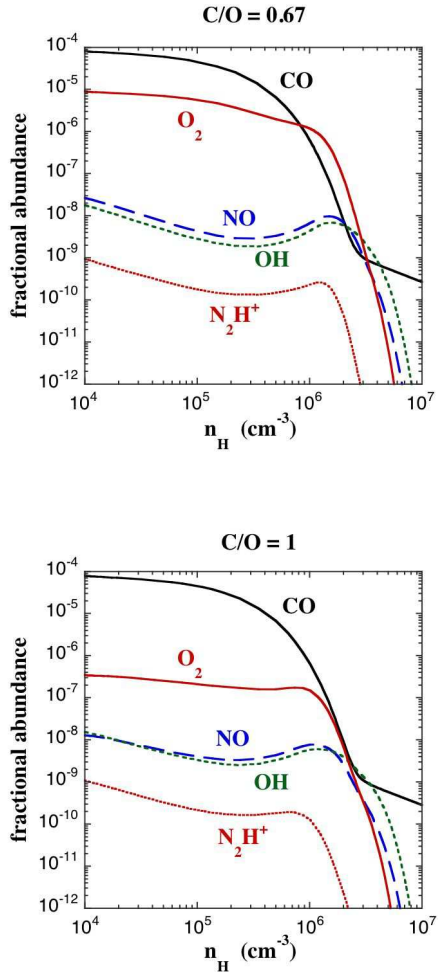}
  \caption{Results of models of the chemistry in a core
	collapsing on the free--fall timescale. Results are
	shown for models in which (a) $n_{\rm C}/n_{\rm O} =
	0.67$ and (b) $n_{\rm C}/n_{\rm O} = 0.97 \approx 1$
	initially. In both cases, the sticking probabilities
	$S({\rm N}) = S({\rm O}) = 0.1$ and $S = 1.0$ for all
	other neutral species.}
  \label{fig_no1}
\end{figure*}

\begin{figure*}
  \centering
  \includegraphics[height=15cm]{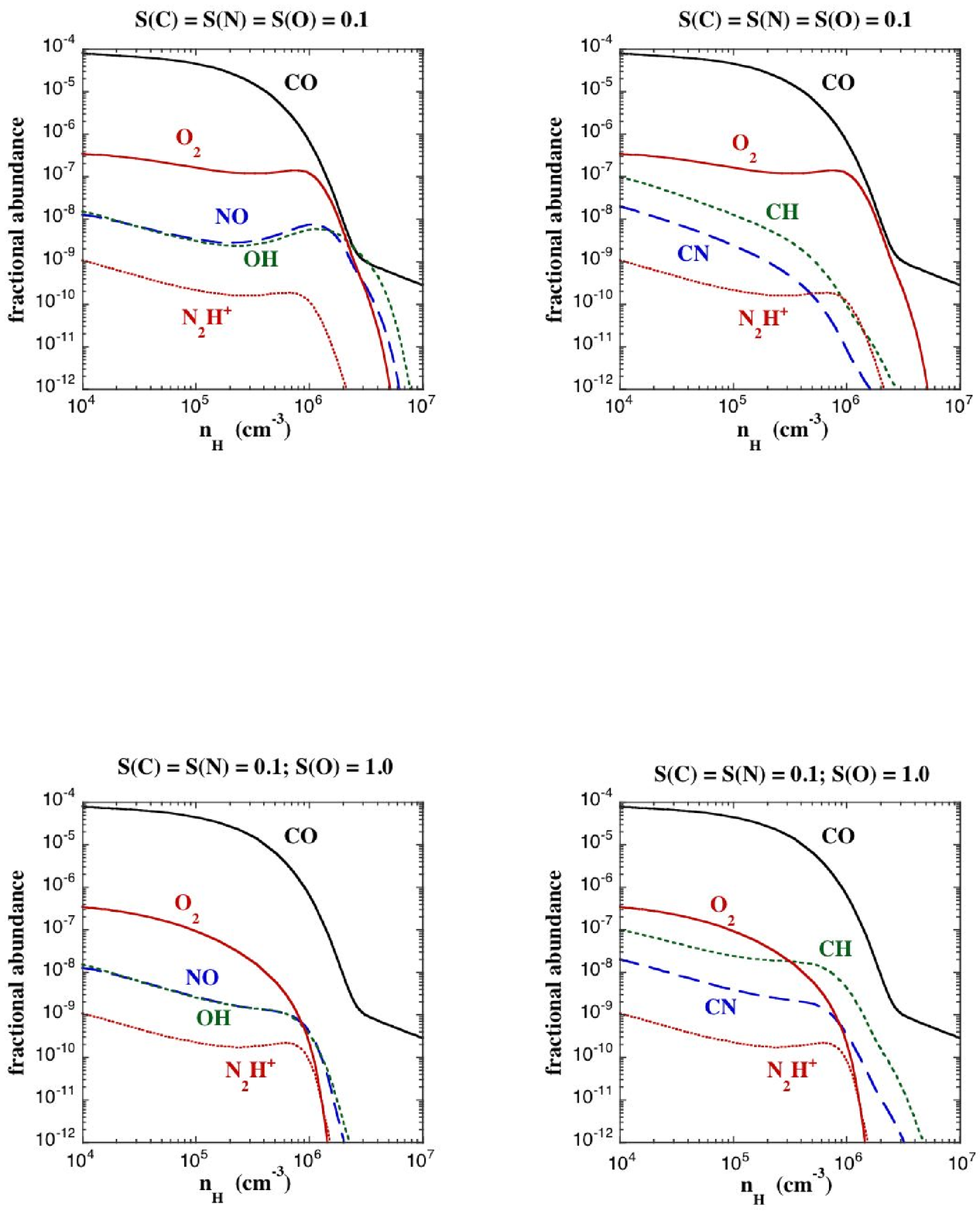}
  \caption{Results of models of the chemistry in a core
	collapsing on the free--fall timescale. Results are
	shown for models in which the grain sticking coefficient
	of atomic oxygen is set equal to $S({\rm O}) = 0.1$
	(upper panels) and $S({\rm O}) = 1.0$ (lower panels). In
	both cases, $S({\rm C}) = S({\rm N}) = 0.1$ and $S =
	1.0$ for all other neutral species; $n_{\rm C}/n_{\rm O}
	= 0.97 \approx 1$ initially in the gas phase.}
  \label{fig_no2}
\end{figure*}

To test some of the above ideas, we have carried out
calculations of the variation of NO, \NTHP , and \AMM \ with
density in the framework of the free--fall collapse model
presented by Flower et al.  (2006).  In this study, it was
found that a good fit to the observations of L1544 in \AMM ,
\NTHP , and their deuterated isotopes required a grain
surface area per H--atom roughly an order of magnitude less
than in the diffuse interstellar medium. These simulations
assumed a single grain size (radius) of typically 0.5 \mic
. It was assumed also that the sticking coefficients for
both atomic nitrogen and oxygen are significantly less than
unity. Then, \AMM \ and \NTHP \ remain in the gas phase at
densities for which CO and other C--containing molecules are
already depleted. In the present Section, we use the model
developed by Flower et al. (2006) and examine its
predictions for the NO abundance. When so doing, we take
account of the fact that our observations indicate that NO
is intermediate between the C--containing species and
molecules like \AMM . The complete chemistry file and list
of chemical species are available from
http://massey.dur.ac.uk/drf/protostellar/species\_chemistry\_08\_06.

One possibility which has been investigated is that our
assumed initial abundances (calculated in steady state for a
density $n_{\rm H} = 10^4$ \percc ) influence the results
significantly.  In particular, we have compared our
`standard model', in which the ratio of elemental carbon to
oxygen in the gas--phase is initially 0.67, to a model in
which this ratio approaches 1. In this latter case, the
underlying assumption is that oxygen which is not initially
in the form of gas--phase CO is in water ice. The results of
these calculations are shown in Fig.~\ref{fig_no1}. We see
from this Figure that the variation of the fractional
abundance of NO, $n({\rm NO})/n_{\rm H}$, with $n_{\rm H}$
is very similar in the two cases; this is because NO forms
in the reaction~(\ref{equ1}) of N with OH. The OH radical is
formed and destroyed in binary reactions involving atomic O,
and so its fractional abundance is insensitive to the total
amount of oxygen in the gas phase. On the other hand, there
is a large decrease in the fractional abundance of O$_2$
when the ratio $n_{\rm C}/n_{\rm O}$ is increased. The
computed fractional abundance of O$_2$ becomes just about
compatible with the upper limit of $n({\rm O}_2)/n({\rm
H}_2) \lesssim 10^{-7}$ derived by Pagani et al. (2003),
from observations of a number of prestellar cores, in the
`carbon--rich' case. For this reason, we adopted $n_{\rm
C}/n_{\rm O} = 0.97 \approx 1$ initially, in the gas phase,
in subsequent models.

We have already mentioned in Sect.~\ref{chemical} that N$_2$
is the precursor of N$_2$H$^+$. Therefore, the initial value
of the fractional abundance of N$_2$H$^+$ depends on the
initial fractional abundance of N$_2$. Because
reactions~(\ref{equ1})--(\ref{equ4}), which transform N into
N$_2$, involve minor species, OH and CH, this transformation
process is slow, requiring approximately $5\times 10^6$~yr
to attain equilibrium at a gas density $n_{\rm H} =
10^4$~cm$^{-3}$. Thus, it is far from clear that the ratio
$n({\rm N})/n({\rm N}_2)$ can be assumed to be initially in
equilibrium (cf. Flower et al. 2006). Accordingly, we have
varied the initial value of this ratio from $n({\rm
N})/n({\rm N}_2) << 1$, obtained assuming that the chemistry
has reached equilibrium, to $n({\rm N})/n({\rm N}_2) >> 1$,
with a view to improving the agreement with the fractional
abundance of \NTHP \ deduced from our observations.

The observed quantities are the column densities of NO,
N$_2$H$^+$ and H$_2$, as functions of position. In order to
compare the free--fall model with the observations, we
relate the offset position to the gas density by means of
the relation $$n(r) = \frac {n(0)}{1+(\frac
{r}{r_2})^{\alpha }}$$ (Tafalla et al.~2002), where $n$ is
the gas density, $r$ is the offset from centre, $r = 0$,
$n(0)$ is the central density of molecular hydrogen, and
$r_2$ is the radial distance over which the density
decreases by a factor of 2, relative to the central
density. From the 1.2~mm dust continuum emission map of
L1544, Tafalla et al. derived $r_2 = 20$\arcsec (equivalent
to 0.014 pc at the distance of L1544), $\alpha = 2.5$, and
$n(0) = 1.4\times 10^6$ cm$^{-3}$. In view of the
freeze--out density indicated by the free-fall model
(Fig.~\ref{fig_no2}), we have adopted a central density
$n(0) = 0.5\times 10^6$ cm$^{-3}$ of molecular hydrogen,
equivalent to $n_{\rm H}(0) = 1.0\times 10^6$ cm$^{-3}$. We
consider that the central density is uncertain to at least
this factor of 2.8 (cf. Bacmann et al. 2000, Crapsi et
al. 2005a, Doty et al. 2005). We note also that the density
$n(r)$ which derives from the above expression is the
maximum value along the line of sight at a given offset,
$r$, from centre.

From our earlier discussion (Sect.~\ref{chemical}), we see
that NO might be sensitive also to the sticking coefficient
of atomic oxygen, $S$(O), and so we compare results for
$S({\rm O}) = 0.1$ and $S({\rm O}) = 1.0$ in
Fig.~\ref{fig_no2}. The results of the calculations are,
indeed, sensitive to $S$(O): NO vanishes from the gas phase
at a density of roughly $10^6$ \percc \ when $S({\rm O}) =
1.0$ but survives to at least 3 times that density when
$S({\rm O}) = 0.1$. Of course, CO is unaffected by the
oxygen sticking probability. In these calculations, $S({\rm
C}) = S({\rm N}) = 0.1$ and $S = 1.0$ for all other neutral
species. The model is `carbon--rich' ($n_{\rm C}/n_{\rm O} =
0.97$ initially in the gas phase), and hence the route to
N$_2$, through the reactions of CH and CN with N
(reactions~(\ref{equ3}) and (\ref{equ4})), assumes
importance, relative to the reactions of OH and NO with N
(reactions~(\ref{equ1}) and (\ref{equ2})). The initial
abundance of CH exceeds that of OH by an order of magnitude.
It is noticeable that CH and CN are considerably more
abundant at densities of the order of 10$^6$ cm$^{-3}$ in
the models in which $S({\rm O}) = 1.0$; atomic oxygen
destroys these species, forming CO. A consequence is that,
if $S({\rm O})$ is large, CN may be expected to be depleted
less than other C--containing species towards the density
peaks of prestellar cores. Finally, we see that the close
link between the fractional abundances of NO and OH,
mentioned in the Introduction and discussed in
Sect.~\ref{chemical}, is apparent in Fig.~\ref{fig_no2}.

\begin{figure*}
  \centering
  \includegraphics[height=20cm]{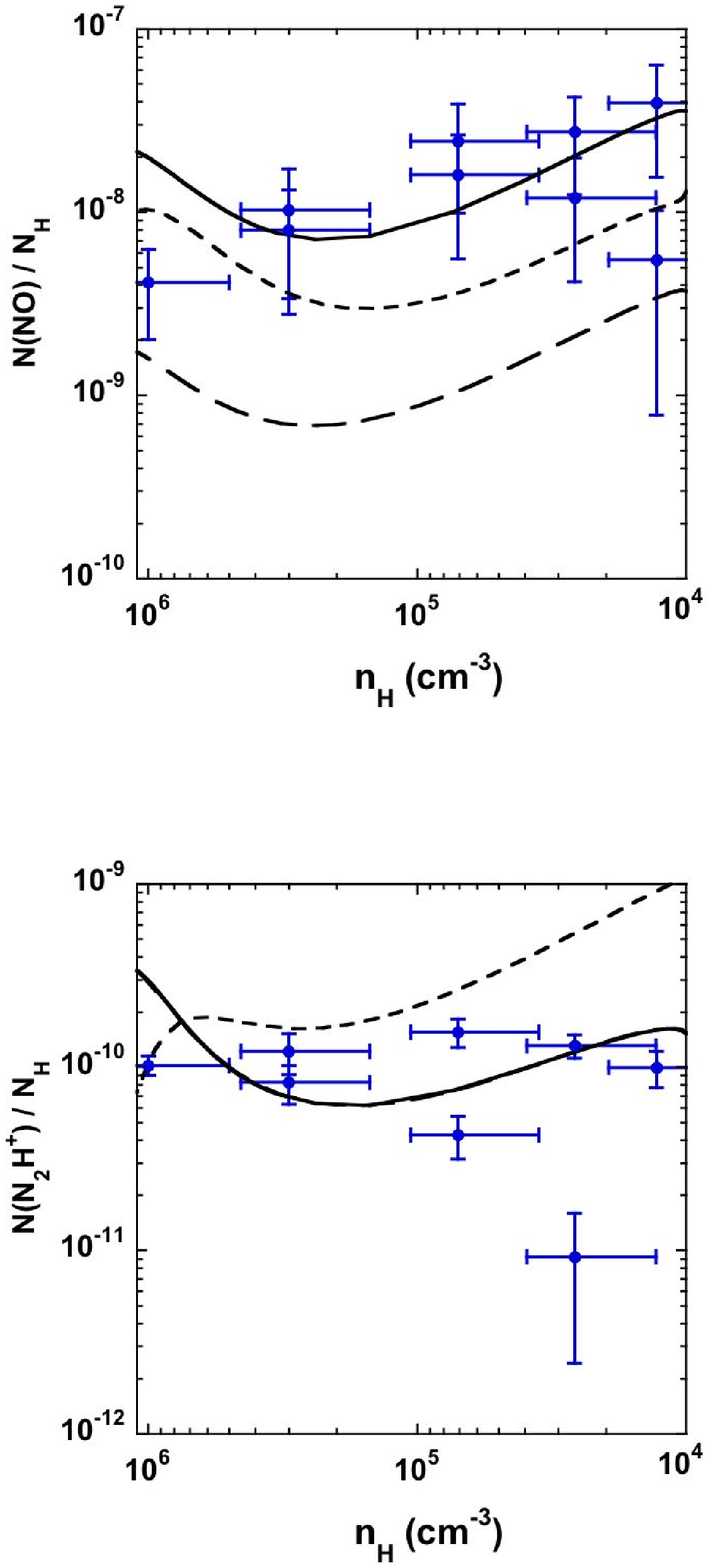}
  \caption{A comparison of the column densities of NO and
	N$_2$H$^+$, relative to N$_{\rm H} \approx 2N({\rm
	H}_2)$, derived from observations along the minor axis
	of L1544 (points with error bars), with the fractional
	abundance of NO, $n({\rm NO})/n_{\rm H}$, and of \NTHP,
	$n({\rm N}_2{\rm H}^+)/n_{\rm H}$, predicted by
	free--fall models in which the initial value of the
	ratio $n({\rm N})/n({\rm N}_2)$ is varied from its
	equilibrium value, $n({\rm N})/n({\rm N}_2) = 1/60$
	(dashed curve) to $n({\rm N})/n({\rm N}_2) = 18$
	(long--dashed curve). The effect of introducing a small
	(26~K) barrier in the rate coefficient of the reaction
	of NO with N (reaction~(\ref{equ2})) is also illustrated
	(full curve); the introduction of the barrier has no
	perceptible effect on \NTHP, where the long--dashed and
	full curves overlap. In these calculations, $S({\rm C})
	= S({\rm N}) = S({\rm O}) = 0.1$ and $S = 1.0$ for all
	other neutral species; initially, $n_{\rm C}/n_{\rm O} =
	0.97 \approx 1$ in the gas phase. Note that the abscissa
	has been reversed to reflect the decrease of the density
	with increasing distance from the centre.}
  \label{fig_no2.1}
\end{figure*}

In Fig.~\ref{fig_no2.1} are compared the observed values of
the column density ratios $N({\rm NO})/N_{\rm H}$ and
$N({\rm N}_2{\rm H}^+)/N_{\rm H}$, where $N_{\rm H} =
2N({\rm H}_2)$, with the computed values of the fractional
abundances $n({\rm NO})/n_{\rm H}$ and $n({\rm N}_2{\rm
H}^+)/n_{\rm H}$. In this Figure, $S({\rm C}) = S({\rm N}) =
S({\rm O}) = 0.1$. The horizontal error bars represent a
factor of 3 uncertainty in $n_{\rm H}$, whilst the vertical
error bars incorporate the observational uncertainties in
the line intensity or flux measurements, together with an
assumed 15~percent calibration error. The observed points
fall on two branches, corresponding to opposite sides of the
central position. The emission from L1544 is observed to be
asymmetric about the centre, whereas the empirical model of
the density distribution, given above, assumes radial
symmetry. (The lower branch corresponds to negative offsets
(in R.A.) for both NO and N$_2$H$^+$.)  It may be seen that
increasing the initial value of the atomic to molecular
nitrogen ratio has the effect of yielding much improved
agreement between the observed and computed profiles of
\NTHP. On the other hand, the profile of NO is displaced
downwards, owing to the enhanced rate of its destruction by
N, and the computed profile now falls well below the
observations. However, if a small barrier introduced into
the rate coefficient for the reaction of NO with N, the
computed NO profile becomes once again compatible with the
observations, as may be seen in Fig.~\ref{fig_no2.1}. In the
calculations presented in this and the following Figures, we
have adopted the rate coefficient for the reaction of NO
with N which is given in the current version of the UMIST
chemical reaction database (LeTeuff et al.~2000), rate05,
namely $k_2 = 3.75\times 10^{-11} {\rm
exp}(-26/T)$~cm$^3$~s$^{-1}$, which derives from the study
of Duff \& Sharma~(1996). This small reaction barrier is
sufficient to inhibit the destruction of NO by N
significantly at $T = 10$~K. We note that it is presently
very difficult to confirm or exclude the existence of such a
small barrier by either laboratory measurements, which
extend to insufficiently low temperatures, or theoretical
calculations, which would need to determine the transition
state energy to an accuracy of 100~$\mu $h.

\begin{figure*}
  \centering
  \includegraphics[height=20cm]{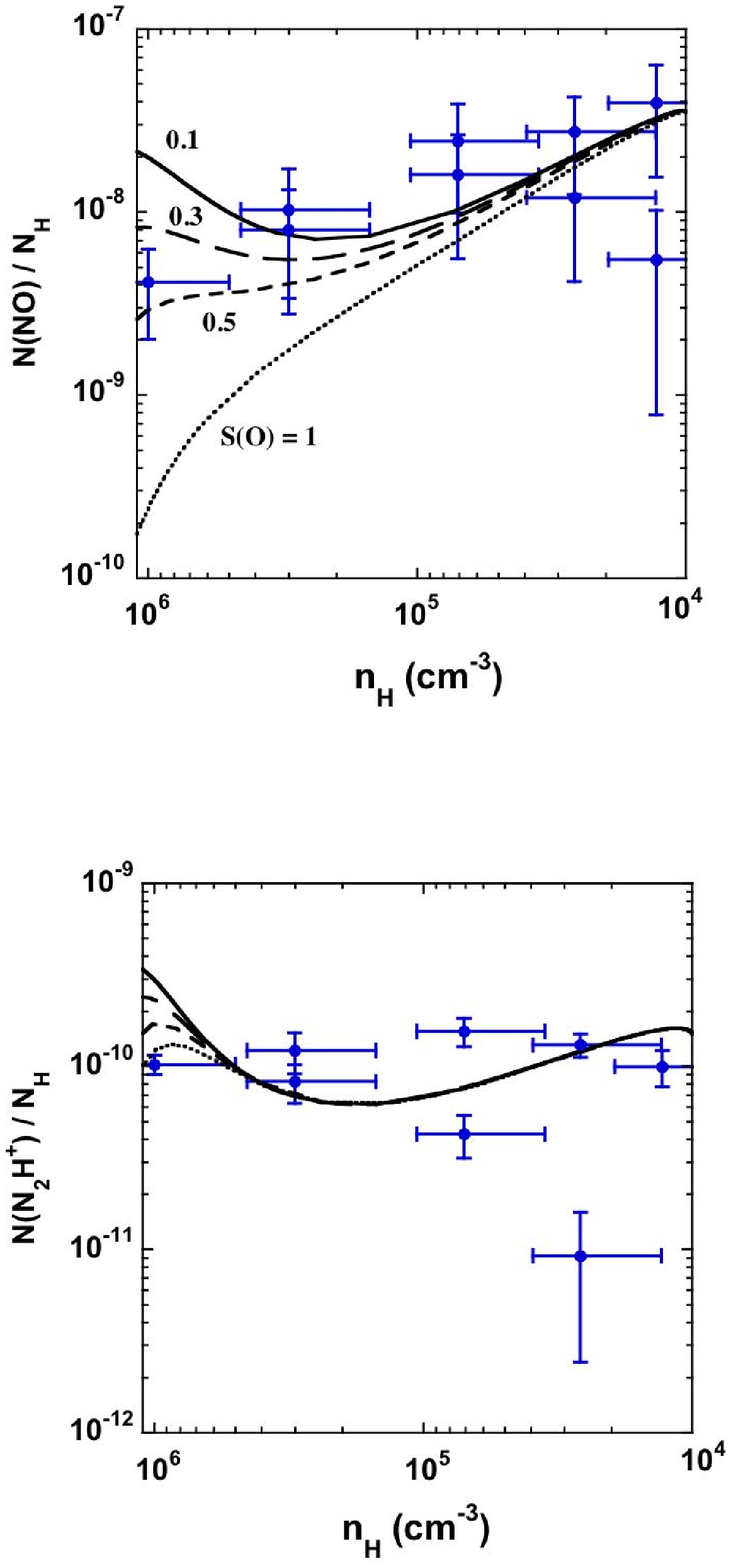}
  \caption{A comparison of the column densities of NO and
	N$_2$H$^+$, relative to N$_{\rm H} \approx 2N({\rm
	H}_2)$, derived from observations along the minor axis
	of L1544 (points with error bars), with the fractional
	abundance of NO, $n({\rm NO})/n_{\rm H}$, and of \NTHP,
	$n({\rm N}_2{\rm H}^+)/n_{\rm H}$, predicted by
	free--fall models in which the grain sticking
	coefficient of atomic oxygen is varied: $S({\rm O}) =
	0.1, 0.3, 0.5, 1.0$. In all cases, $S({\rm C}) = S({\rm
	N}) = 0.1$ and $S = 1.0$ for all other neutral species;
	initially, $n_{\rm C}/n_{\rm O} = 0.97 \approx 1$ in the
	gas phase, and $n({\rm N})/n({\rm N}_2) = 18$. Note that
	the abscissa has been reversed to reflect the decrease
	of the density with increasing distance from the
	centre.}
  \label{fig_no3}
\end{figure*}

In Fig.~\ref{fig_no3}, we investigate the effect of varying
the grain sticking coefficient of atomic oxygen, in the
range in the range $0.1 \le S({\rm O}) \le 1.0$. Once again,
$S({\rm C}) = S({\rm N}) = 0.1$.  We see that the general
trends of the observations are for $N({\rm NO})/N_{\rm H}$
to increase with decreasing density (increasing distance
from the centre), whereas $N({\rm N}_2{\rm H}^+)/N_{\rm H}$
remains approximately constant or, possibly, tends to
decrease with decreasing density; this is in accord with the
comments made in the previous Section. The models in which
$S({\rm O}) > 0.1$ predict that the fractional abundance of
N$_2$H$^+$ is roughly constant, which, given the large error
bars, is in reasonable accord with the observations. As
$S({\rm O})$ increases towards unity, the amount of oxygen
remaining in the gas phase at high densities decreases and
the rate of formation of NO, in reactions~(\ref{equ1}) and
(\ref{equ2}), decreases also. When $S({\rm O}) = 0.5$,
$n({\rm NO})/n_{\rm H} = 3.5\times 10^{-8}$ when $n_{\rm H}
= 10^{4}$~cm$^{-3}$, and $n({\rm NO})/n_{\rm H} = 2.9\times
10^{-9}$ when $n_{\rm H} = 10^{6}$~cm$^{-3}$; these values
are consistent with those deduced from the empirical model
(Sect.~\ref{empirical}). However, we wish to emphasize that
we do not consider ourselves in a position to ``determine''
the sticking coefficients: this should be done in the
laboratory.  We have treated the sticking coefficients of C,
N, and O simply as parameters of the model, varying their
values in an attempt at fitting the available observations.

In our recent study (Flower et al. 2006), we found that
satisfactory agreement could be obtained with the observed
high levels of deuteration of the nitrogen--containing
species NH$_3$ and \NTHP, providing that the grains were
sufficiently large ($a_{\rm g} \approx 0.50$~$\mu $m, as in
the present calculations) and that the sticking
probabilities of atomic nitrogen and atomic oxygen were both
less than unity ($S \lesssim 0.3$). For the reasons given
above, we have since modified the initial values of the C/O
and N/N$_2$ ratios, as well as reducing the sticking
probability of atomic carbon to $S({\rm C}) = 0.1$. We have
seen (Fig.~\ref{fig_no3}) that the observed variations of
the column densities of NO and \NTHP \ can then be
reproduced, but it is necessary to establish that this
success is not at the expense of the earlier agreement with
the degrees of deuteration of NH$_3$ and \NTHP.

\begin{figure*}
  \centering
  \includegraphics[height=20cm]{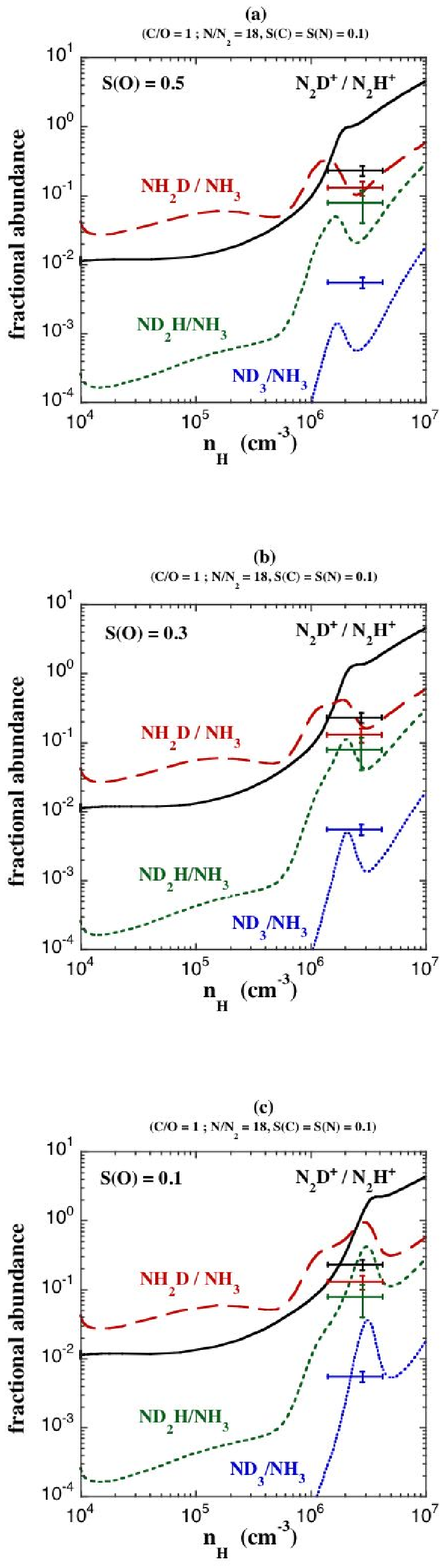}
  \caption{The abundance ratios of the deuterated forms of
	N$_2$H$^+$ and NH$_3$, as functions of $n_{\rm H}$,
	together with the values observed in the prestellar core
	L1544 (cf. Flower et al. 2006, figure~5c). The grain
	sticking probability of atomic oxygen, $S({\rm O}) =
	0.5, 0.3, 0.1$ in panels (a), (b), (c), respectively. In
	all cases, $S({\rm C}) = S({\rm N}) = 0.1$ and $S = 1.0$
	for all other neutral species; initially, $n_{\rm
	C}/n_{\rm O} = 0.97 \approx 1$ in the gas phase, and
	$n({\rm N})/n({\rm N}_2) = 18$.}
  \label{fig_no4}
\end{figure*}

In Fig.~\ref{fig_no4} are plotted the relative abundances of
the deuterated and non-deuterated forms of NH$_3$ and \NTHP
\ which derive from the present calculations, adopting
$n_{\rm C}/n_{\rm O} = 0.97$ and $n({\rm N})/n({\rm N}_2) =
18$ initially, together with $S({\rm C}) = S({\rm N}) =
0.1$; the sticking probability of atomic oxygen is varied in
the range $0.1 \le S({\rm O}) \le 0.5$. The observational
points and their error bars are identical to those adopted
in our earlier paper (Flower et a. 2006, figure~5c). It may
be seen from Fig.~\ref{fig_no4} that satisfactory agreement
with the observed levels of deuteration is still obtained
when $S({\rm O}) \approx 0.3$. This Figure emphasizes the
fact that the degree of deuteration of NH$_3$ increases with
increasing gas density. The fractional abundance of ND$_3$,
for example, has a distinct maximum at densities of order
$10^6$~cm$^{-3}$. As kinematical parameters, such as the
collapse speed, might be expected to vary with radial
distance (and hence density) in the core, the widths of
emission lines from successive stages of deuteration of
NH$_3$ could provide information on the variation of the
infall speed with the gas density.

We note that the effect of subsequent coagulation on the
size of initially large grains -- specifically, $a_{\rm g} =
0.50$~$\mu $m in the present models -- is small, over the
range of compression considered ($10^4 \le n_{\rm H} \le
10^7$~cm$^{-3}$), and has little consequence for the
freeze--out timescale.

\section{Concluding remarks}

Our principal observational conclusion is that NO, unlike
\AMM \ and \NTHP, is depleted towards the density peaks of
the prestellar cores L1544 and L183. This result was
surprising, in that NO was believed to be an important
intermediate of the nitrogen chemistry. In particular, the
path from atomic to molecular nitrogen was believed to
involve both OH and NO. Our results cast some doubt on this
assumption, suggesting that molecular nitrogen may form from
CH and CN at high densities. If so, CN is an important
intermediate in the nitrogen chemistry. Further mapping of
CN in prestellar cores, extending the observations of L183
by Dickens et al. (2000), would be extremely useful. Our
models suggest a fractional abundance $n({\rm CN})/n_{\rm H}
\approx 10^{-10}$ towards the peak of L1544, corresponding
to a column density of the order of $10^{13}$~cm$^{-2}$. In
the case of L183, the central depression in the fractional
abundance of NO is offset from the dust peak. This object
needs further study: complete maps of the NO distribution
are required.

In an attempt to explain our observations within the
framework of a free--fall collapse model, we have varied the
values of the sticking probabilities of atomic C, N and O,
and also the initial values of the elemental C/O and the
atomic to molecular nitrogen abundance ratios in the gas
phase. We find that

\begin{itemize}
  
\item the prestellar core L1544 is likely to be
  `carbon--rich', in the sense that $n_{\rm C}/n_{\rm O}
  \approx 1$ in the gas phase. An initial $n_{\rm C}/n_{\rm
  O} = 0.97$ leads to values of $n({\rm O}_2)/n({\rm H}_2)$
  which are in much better agreement with the observed upper
  limits than when the gas is oxygen--rich ($n_{\rm
  C}/n_{\rm O} = 0.67$).

In objects in which the gas--phase C/O abundance ratio is
close to unity, the fractional abundances of C--rich
species, such as the cyanopolynes, are expected to be
enhanced relative to oxygen--containing species, such as
O$_2$ and H$_2$O. Of the two cores which we have studied,
this is true of L1544 (cf. Walmsley et al. 1980), whereas
L183 is relatively carbon--poor (cf. Swade 1989).

\item The ratio $n({\rm N})/n({\rm N}_2)$ at the start of
  collapse is much more than its value in equilibrium. This
  result is perhaps not surprising, given the long timescale
  ($\approx 5\times 10^6$~yr) required for the nitrogen
  chemistry to reach equilibrium at densities $n_{\rm H}
  \approx 10^4$~cm$^{-3}$. We note that Maret et al. (2006)
  have recently and independently arrived at essentially the
  same conclusion, viz. that most of the nitrogen in the gas
  phase is atomic, also from an analysis of observations of
  \NTHP, in their case in the prestellar core B68 (but with
  implications for the deficiency of molecular nitrogen in
  cometary ices). If $n({\rm N})/n({\rm N}_2) >> 1$ in the
  gas phase, then our observations of NO imply the existence
  of a small barrier ($\approx 25$~K) to the reaction of NO
  with N (equation~(\ref{equ2})). Experimental demonstration
  of the existence of such a small barrier to this and,
  possibly, the other key neutral--neutral
  reactions~(\ref{equ1}), (\ref{equ3}) and (\ref{equ4}),
  whilst difficult to achieve, is highly desirable.

\item The grain sticking probabilities of atomic C, N and,
  probably, O are significantly smaller than unity; values
  of $S({\rm C}) = S({\rm N}) = 0.1$ and $0.1 \le S({\rm O})
  \le 1.0$ were adopted in the calculations reported
  above. However, the assumed values of the sticking
  probabilities for the atomic species are ad hoc and
  require confirmation by means of experimental
  measurements. Such measurements should presumably be made
  on CO ice, as in the experiments of Bisschop et
  al. (2006).

\end{itemize}

\appendix

\section{Determination of the column densities}

The method for deriving the column density of NO, from
observations of the hyperfine components of rotational
transitions, assumed optically thin, has been presented by
Gerin et al.~(1992); we followed their method in the present
paper. The case of \NTHP \ was considered by Caselli et
al.~(2002), who gave expressions for the total column
density of \NTHP, for the case of a finite optical depth in
the line and also in the optically thin limit, in their
Appendix A. If the optical depth in a line of central
frequency $\nu $ and wavelength $\lambda $ is finite, then
the {\it total} column density of the molecular ion is
related to to the optical depth $\tau $ at the line centre
by

\begin{equation}
  N = \frac {4\pi \Delta v}{\lambda ^3} \left (\frac
  {\pi}{{\rm ln}\,2} \right )^{\frac {1}{2}} \frac
  {Q}{g_{\rm u} A({\rm u} \rightarrow {\rm l})} \frac {\tau
  }{1-\exp(-h\nu /k_{\rm B}T_{\rm ex})} \frac
  {1}{\exp(-E_{\rm l}/k_{\rm B}T_{\rm ex})}
\label{equA1}
\end{equation}
where $\Delta v$ is the full width of the line at
half-maximum intensity (the line profile is assumed to be
gaussian), $g_{\rm u}$ is the degeneracy (statistical
weight) of the upper level of the observed transition,
$A({\rm u} \rightarrow {\rm l})$ is the spontaneous
radiative transition probability to the lower level, whose
excitation energy is $E_{\rm l}$; $Q$ is the partition
function, considered below.  We have assumed implicitly a
Boltzmann distribution of population at an excitation
temperature (to be determined) $T_{\rm ex}$. In the limit of
low optical depth at the line centre, the expression for the
total column density is

\begin{displaymath}
  N = \frac {4\pi \Delta v}{\lambda ^3} \left (\frac
  {\pi}{{\rm ln}\,2} \right )^{\frac {1}{2}} \frac
  {Q}{g_{\rm u} A({\rm u} \rightarrow {\rm l})} \frac
  {(k_{\rm B} T_{\rm mb}/h \nu )}{[\exp (h\nu /k_{\rm
  B}T_{\rm ex}) - 1]^{-1} - [\exp (h\nu /k_{\rm B}T_{\rm
  bg}) - 1]^{-1}}
\end{displaymath}
\begin{equation}
  \times \frac {1}{1-\exp(-h\nu /k_{\rm B}T_{\rm ex})} \frac {1}{\exp(-E_{\rm l}/k_{\rm B}T_{\rm ex})}
  \label{equA2}
\end{equation}
where $T_{\rm mb}$ is the main beam brightness temperature
and $T_{\rm bg}$ is the temperature of the background
radiation field, usually taken to be a black--body at
$T_{\rm bg} = 2.73$~K.

The line width, $\Delta v$, and the brightness temperature,
$T_{\rm mb}$, refer to a given component of the $J = 1
\rightarrow 0$ multiplet, which is observed towards L1544
and L183 to comprise seven resolved components. The full set
of transitions is listed in Table~\ref{trans}, along with
the corresponding line strengths. A comparison of the
relative intensities of the components with the values
expected in the optically thin limit (when they are given by
the line strengths) shows that some of the lines are
optically thick towards the dust emission peaks of these two
sources. Accordingly, we have used the weakest transition,
$J, F_1, F = 1, 1, 0 \rightarrow 0, 1, 1$ of the multiplet,
which should also be the most optically thin, to determine
the column density of \NTHP. Previous work has shown that
\NTHP \ column densities derived in this way agree to within
about 30 percent with the results of more sophisticated
analyses of the emission in the rotational multiplet, which
allow for optical depth effects; this is the case even at
the peak of the emission, where the assumption that the $1,
1, 0 \rightarrow 0, 1, 1$ transition is optically thin is
least valid. Crapsi et al.~(2005a) estimated the``total''
optical depth (i.e. summed over all seven components) of the
N$_2$H$^+$ $J = 1 \rightarrow 0$ multiplet to be $\tau =
12.6\pm 0.7$ towards the peak of L1544, which corresponds to
an optical depth in the $1, 1, 0 \rightarrow 0, 1, 1$
transition of $12.6/27 = 0.47$ and an escape probability of
0.80 (and hence a 25 percent underestimation of the column
density owing to the ``optically thin'' assumption). They
estimated, also from their fit, an excitation temperature
$T_{\rm ex} = 5$~K for the $J = 1 \rightarrow 0$ multiplet
in \NTHP.

The hyperfine splitting of the rotational energy levels of
$^{14}$N$_2$H$^+$ arises from the interaction of the spin
angular momentum, ${\bf I}$, of the nitrogen nuclei and the
rotational angular momentum, ${\bf J}$, of the molecule. The
angular momentum coupling scheme is

\begin{displaymath}
  {\bf F_1} = {\bf J} + {\bf I_1}
\end{displaymath}
and

\begin{displaymath}
  {\bf F} = {\bf F_1} + {\bf I_2}
\end{displaymath}
The nuclear spin angular momentum quantum number is $I_1 =
I_2 = 1$. When $J \ge 1$, there are three possible values of
$F_1$, namely $F_1 = J - 1, J, J + 1$. On the other hand,
when $J = 0$, $F_1 = I_1 = 1$. The possible values of $F$,
the total angular momentum, are determined by the (weaker)
coupling to the spin of the second (inner) nitrogen
nucleus. The quantum numbers $J, F_1, F$ which identify the
hyperfine levels of the $J = 1$ and $J = 0$ rotational
states are listed in Table~\ref{trans}. The optically
allowed transitions are determined by the electric dipole
selection rules $\Delta F = 0, \pm 1$ (but $\Delta F = 0$,
when $F = 0$, is not allowed) governing the change in $F$ in
the transition. The hyperfine splitting of $J = 0$, which
arises from the nuclear spin--spin interaction, is very
small: see Green et al. (1974), Caselli et
al. (1995). Consequently, allowed transitions from a given
upper level, $J = 1, F_1, F$, are unresolved, and there
remain seven resolved hyperfine transitions between $J = 1$
and $J = 0$, one from each of the hyperfine levels of $J =
1$.

The partition function, $Q$, which appears in equation~(\ref{equA2}), is defined by

\begin{equation}
Q = \sum _i g_i \exp \left (-\frac {E_i}{k_{\rm B} T_{\rm ex}}\right )
\label{equA4}
\end{equation}
where the summation extends over all energy levels, and $g_i = 2F_i + 1$. As $E_i$ is approximately the same for all hyperfine levels belonging to the same $J$, the summation over $g_i$ for each value of $J$ may be carried out, yielding

\begin{equation}
Q = 9 \sum _i g_J \exp \left (-\frac {E_J}{k_{\rm B} T_{\rm ex}}\right )
\label{equA5}
\end{equation}
where $g_J = 2J + 1$. The factor of 9 appearing in
(\ref{equA5}) is the product of the degeneracies, $2I + 1 =
3$, arising from the spin of the nitrogen nuclei. It may be
verified that, for a given value of $J$,

\begin{equation}
\sum _i (2F_i + 1) = 9\, (2J + 1)
\label{equA6}
\end{equation}

Assuming that the hyperfine splitting is small compared with
the rotational splitting of the energy levels (which is
verified, in practice), the relationship between the
probability of a given hyperfine transition within a
rotational multiplet and the total probability of a
transition between the rotational levels $J^{\prime}$, $J$
is

\begin{equation}
g_{F^{\prime}} A(J^{\prime} F_1^{\prime} F^{\prime} \rightarrow J F_1 F) = 9\, g_{J^{\prime}} A(J^{\prime} \rightarrow J) S(F_1^{\prime} F^{\prime}, F_1 F)
\label{equA3}
\end{equation}
where $g_{F^{\prime}} = 2F^{\prime} + 1$ and $g_{J^{\prime}}
= 2J^{\prime} + 1$; $9\, g_{J^{\prime}}$ is the total
degeneracy of rotational state $J^{\prime}$
(cf. equation~(\ref{equA6})). $S(F_1^{\prime} F^{\prime},
F_1 F)$ is the line strength, which determines the relative
intensities of the hyperfine components of the rotational
multiplet. The sum of the line strengths, over all
components of the multiplet, is equal to unity. Because the
hyperfine levels of $J = 0$ are almost degenerate, the line
strengths may be summed over over all allowed transitions
from a given hyperfine level of $J = 1$. In
Table~\ref{trans}, the resulting line strengths are listed
for the seven resolved transitions from $J^{\prime} = 1$ to
$J = 0$ in \NTHP; they may be seen to be given by the ratio
$(2F^{\prime} + 1)/\sum (2F^{\prime} + 1)$, where the
summation extends over the hyperfine levels of the upper
rotational state, $J^{\prime}$. When $J^{\prime} = 1$, $\sum
(2F^{\prime} + 1) = 27$, as may be seen from
equation~(\ref{equA6}); see \cite{townes55} and
\cite{tatum86} for more detailed discussions of coupling
schemes and line strengths. The specific case of N$_2$H$^+$
is considered in a recent publication of Daniel et
al. (2006) (but note that $J^{\prime }$ in their
equation~(1) for the transition probability should read
$J$).

\begin{table}
\caption{The $J^{\prime} = 1 \rightarrow J = 0$ hyperfine
  transitions at 93.2~GHz. The seven resolved lines are
  ordered in increasing frequency, from bottom to top. For
  each transition, the quantum numbers identifying the upper
  level (with primes) and lower level are given, along with
  the corresponding line strength. The line strengths have
  been summed over the allowed transitions from a given
  hyperfine level of $J = 1$, given that the hyperfine
  levels of $J = 0$ are almost degenerate.}
\vspace{1em}
\begin{tabular}{cc}
\hline
\hline
    Transition  & Line strength  \\
    $J^{\prime} F_1^{\prime} F^{\prime} \rightarrow J F_1 F$  & $S(F_1^{\prime} F^{\prime}, F_1 F)$  \\
\hline
     $1 0 1 \rightarrow 0 1 2$ & $1/9$  \\
     $1 2 1 \rightarrow 0 1 1$ & $1/9$  \\
     $1 2 3 \rightarrow 0 1 2$ & $7/27$ \\
     $1 2 2 \rightarrow 0 1 1$ & $5/27$ \\
     $1 1 1 \rightarrow 0 1 0$ & $1/9$  \\
     $1 1 2 \rightarrow 0 1 2$ & $5/27$ \\
     $1 1 0 \rightarrow 0 1 1$ & $1/27$ \\
\hline
\hline
\end{tabular}
\label{trans}
\end{table}

The derivation of the total column density of a molecule
from observations of individual transitions -- hyperfine
transitions between rotational states in the cases of NO and
\NTHP \ -- relies on correcting for the populations of
unobserved levels. When this is done via
equation~(\ref{equA4}) for the partition function, $Q$, it
is assumed that the relative level populations may be
characterized by a Boltzmann distribution at the excitation
temperature, $T_{\rm ex}$; this is the case when collisional
de-excitation of the upper to the lower rotational state
dominates spontaneous radiative decay. The `critical
density' is defined as the density of the collisional
perturber (predominantly H$_2$ in the case of rotational
transitions) for which the radiative and collisional decay
rates become equal (once a correction has been made for the
photon escape probability). For perturber densities higher
than the critical density, a Boltzmann distribution of the
relative level populations is approached, at the kinetic
temperature of the gas. Corrections can be made for optical
depth effects by means of either ``large velocity gradient''
programs (e.g. Crapsi et al.~2004) or more sophisticated
(Monte--Carlo) radiative transport models (e.g. Tafalla et
al.~2002).

The peak column density of \NTHP \ in L1544 derived by
Crapsi et al.~(2005a), $N$(N$_2$H$^+$) $= (1.83 \pm
0.19)\times 10^{13}$~cm$^{-2}$, who used the weakest
component and the optically thin approximation, agrees to
within the error bar with the determination of Caselli et
al. (2002), $N$(N$_2$H$^+$) $= 2.0\times 10^{13}$~cm$^{-2}$,
who used all seven hyperfine components and corrected for
the finite optical depths. Thus, a simple approach to the
determination of the column density of \NTHP \ can yield
satisfactory results. To our knowledge, all line transfer
models to date have assumed that the density structure of
L1544 can be represented by a one--dimensional, spherically
symmetric distribution. It would be interesting to compare
with the predictions of radiative line transfer models which
go beyond the approximation of a single dimension, even
though the angular resolution and completeness of the
currently available maps of the molecular emission would
constrain the comparison of such models with the
observations.

We are not aware of calculations of rate coefficients for
rotational transitions in NO, induced by H$_2$. However,
calculations have been performed for OH -- H$_2$ (Offer \&
van Dishoeck~1992), which shares many physical
characteristics with NO -- H$_2$, not least in the
rotational structure of the target molecule. In the case of
collisions with para--H$_2$, which is the main form of H$_2$
at low temperatures, the calculations of Offer \& van
Dishoeck (1992) indicate that the rate coefficient for
rotational de-excitation at $T = 50$~K is of the order of
$10^{-11}$ cm$^{3}$ s$^{-1}$ and not strongly dependent on
the kinetic temperature, $T$. The rate of spontaneous
radiative decay in the transitions of NO which we observed
(cf. Table~\ref{nofreq}) is of the order of $10^{-7}$
s$^{-1}$ (Gerin et al. 1992), which implies a critical
density of H$_2$ of the order of $10^4$ cm$^{-3}$. In the
regions of the protostellar cores considered above, $n({\rm
H}_2) \gtrsim 10^4$ cm$^{-3}$, extending up to values in
excess of $10^6$ cm$^{-3}$. We conclude that the assumption
of Boltzmann equilibrium at the local kinetic temperature is
justified in the case of NO.

Consider now \NTHP. Daniel et al. (2005) have recently
computed rate coefficients for the rotational de-excitation
of \NTHP \ by He; in the absence of data specifically for
H$_2$, we adopt their results as a guide. Their calculations
indicate a value of $10^{-11}$ cm$^{3}$ s$^{-1}$ for
rotational transitions in \NTHP, induced by He, at low
temperatures, similar to the value above for NO -- H$_2$. On
the other hand, the rate of spontaneous decay of $J = 1$ to
$J = 0$ is of the order of $10^{-5}$ s$^{-1}$ (Dickinson \&
Flower 1981), 2 orders of magnitude larger than for NO
(\NTHP \ has a much larger dipole moment). Consequently, the
critical density for \NTHP \ is of the order of $10^6$
cm$^{-3}$, although line trapping will tend to reduce this
value towards the peaks of L1544 and L183.

\begin{acknowledgements}

GdesF and DRF gratefully acknowledge support from the
`Alliance' programme, in 2004 and 2005.  CMW and MA
acknowledge support from RADIONET for travel to the 30~m
telescope. We are grateful to Paola Caselli for her comments
on an earlier version of this paper, to Ian Sims for helpful
advice, and to an anonymous referee for constructive
suggestions.

\end{acknowledgements}

\end{document}